The Journal of Chemical Physics

RESEARCH ARTICLE | APRIL 22 2024
# Effect of the presence of pinned particles on the structural parameters of a liquid and correlation between structure and dynamics at the local level

Palak Patel; Sarika Maitra Bhattacharyya

# Effect of the presence of pinned particles on the structural parameters of a liquid and correlation between structure and dynamics at the local level

Palak Patel[1,2] 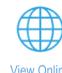 and Sarika Maitra Bhattacharyya[1,2,a)]

**AFFILIATIONS**

[1] Polymer Science and Engineering Division, CSIR-National Chemical Laboratory, Pune 411008, India
[2] Academy of Scientific and Innovative Research (AcSIR), Ghaziabad 201002, India

a)Author to whom correspondence should be addressed: mb.sarika@ncl.res.in

**ABSTRACT**

Pinning particles at the equilibrium configuration of the liquid is expected not to affect the structure and any property that depends on the structure while slowing down the dynamics. This leads to a breakdown of the structure dynamics correlation. Here, we calculate two structural quantities: the pair excess entropy, $S_2$, and the mean field caging potential, the inverse of which is our structural order parameter (SOP). We show that when the pinned particles are treated the same way as the mobile particles, both $S_2$ and SOP of the mobile particles remain the same as those of the unpinned system, and the structure dynamics correlation decreases with an increase in pinning density, "c." However, when we treat the pinned particles as a different species, even if we consider that the structure does not change, the expression of $S_2$ and SOP changes. The microscopic expressions show that the interaction between a pinned particle and a mobile particle affects $S_2$ and SOP more than the interaction between two mobile particles. We show that a similar effect is also present in the calculation of the excess entropy and is the primary reason for the well-known vanishing of the configurational entropy at high temperatures. We further show that, contrary to the common belief, the pinning process does change the structure. When these two effects are considered, both $S_2$ and SOP decrease with an increase in "c," and the correlation between the structural parameters and the dynamics continues even for higher values of "c."

## I. INTRODUCTION

When a glass forming liquid is cooled fast enough, it avoids the crystallization process, and the viscosity/relaxation timescale shows a dramatic increase.[1,2] There have been debates about the origin of this increase in viscosity/relaxation time. There are theories suggesting that the slowing down of the dynamics is purely kinetic in nature.[3] However, phenomenological Adam–Gibbs (AG) theory predicts a relation between the relaxation time, $\tau$, and configurational entropy, $S_c$, $\tau = \tau_0 \exp \frac{A}{TS_c}$, where $A$ is a system dependent constant and $\tau_o$ is the high temperature relaxation time. As predicted by Kauzman many years ago, $S_c$ vanishes at the Kauzmann temperature, $T_K$, which is a finite temperature below the glass transition temperature.[4] For many systems, the AG relation is found to be valid, and the predicted temperature where the dynamics diverges is found to be the same as $T_K$.[5–13] The random first-order transition theory (RFOT) suggests that the slowing down of the dynamics is related to a growing length scale in the system,[14–16] which in turn is related to the configurational entropy, $S_c$, of the system, thus suggesting a generalized AG relationship.[17,18] However, the ideal glass transition temperature $T_K$ can be obtained only by extrapolating the temperature dependence of $S_c$ to low temperatures.

In order to access the ideal glass transition temperature, $T_K$, a novel model system was proposed where some fraction of particles in their equilibrium liquid configuration are pinned.[19–23] It was predicted[19] and also shown in numerical simulations[20,21] that as the fraction of pinned particles increases, the $T_K$ increases, and eventually, at high enough pinning, the ideal glass transition moves to a high enough temperature where the system can be equilibrated. Interestingly, the pinned system can also be experimentally realized by laser pinning some particles[24] or via soft pinning.[25]

Studies showed that for these pinned systems, although the configurational entropy vanishes at high temperatures, the dynamics continues, and there is a breakdown of the AG relationship.[20,22,26] It was later shown that in the calculation of the vibrational entropy, when anaharmonic contributions are considered, the configurational entropy remains positive.[27] However, even with this anharmonic contribution, the AG relationship was shown to break down.[22] It was also shown that the RFOT theory, which leads to a generalized AG relationship, is valid if it is assumed that the configurational entropy of the pinned system is related to the unpinned system by a multiplicative factor where the factor decreases with increasing pinning.[21,28] All these studies showed that the correlation between the dynamics and entropy of the pinned system differs from that of the unpinned system.

The correlation between local pair excess entropy, which depends on the structure and the local dynamics of the pinned system, was also studied.[24] It was shown that since the pinning process is expected not to affect the structure, the local pair excess entropy remains the same as in the unpinned system, whereas the dynamics slows down due to pinning. Thus, there is a decorrelation between pair excess entropy and dynamics, even at the microscopic level.

From the above discussion, it appears that both at macroscopic and microscopic levels, the dynamics and the entropy are not correlated. However, at the macroscopic level, pinning decreases the configurational entropy more than slowing down the dynamics,[20] whereas at the microscopic level, the pinning process does not alter the pair excess entropy but slows down the dynamics. Thus, the decorrelation between entropy and dynamics observed at the macroscopic and microscopic levels is just the opposite. Note that for the unpinned system, the macroscopic pair excess entropy, $S_2$, contributes to 80% of the excess entropy.[29] The configurational entropy $S_c = S_{id} + S_{ex} - S_{vib}$ has a contribution from three terms: the ideal gas entropy $S_{id}$, the excess entropy $S_{ex}$, and the vibrational entropy $S_{vib}$. Since pair excess entropy does not change due to pinning, we can expect the excess entropy, which is usually obtained using the thermodynamic integration (TI) method,[30,31] also not to change. In that case, we may expect that the other terms are responsible for the observed decrease in the configurational entropy of the pinned systems.

In this paper, we first revisit the calculation of the configurational entropy. We show that the decrease in the excess entropy is primarily responsible for the decrease in the configurational entropy. We further show that in the calculation of the excess entropy, the pinned particles should be treated as a different species, and we should consider that the total excess entropy is only calculated for the mobile particles. Under these assumptions, the analytical expression predicts that, compared to a mobile particle, a pinned particle has a stronger contribution in lowering the entropy of mobile particles. We next show that when we use a similar methodology in the calculation of the pair excess entropy, both at macroscopic and microscopic levels, it decreases with pinning. The expression of the pair excess entropy shows that this decrease again comes from the stronger effect of the pinned particles on the mobile particles.

We then extend the recently developed theoretical formulation, where we describe each particle as having a mean field caging potential for the pinned system. Note that, as shown before, this mean field caging potential is obtained from the structure of the liquid.[32–34] We find that even the mean field potential, both at microscopic and macroscopic levels, shows that the pinned particles have a stronger interaction with the mobile particles, thus increasing the depth of the caging potential and confining the mobile particles. We refer to the inverse depth of the caging potential as the structural order parameter (SOP). Interestingly, a similar confinement effect of the pinned particles was observed in the elastically collective nonlinear equation (ECNLE) theory.[35,36] In the ECNLE theory, the dynamics of the system was obtained using Langevin dynamics on the dynamic free energy surface. The studies showed that with pinning, the depth of the free energy barrier increases, and the particles are more confined. In the ECNLE theory, while treating the system, the authors considered that the pinned particles do not change the structure, but the pinned particles are considered to be a different species. Thus, it appears that in any formulation to obtain the stronger confinement effect by the pinned particles, the pinned particles should be treated as a different species.

We next show that contrary to common belief, the pinning process does change the structure, which can be observed only when the partial radial distribution functions are calculated, assuming the pinned particles are of a different species. Our study reveals that with an increase in pinning density, it is a combined effect of the change in structure and the stronger contribution of pinned particles in decreasing the potential energy of the mobile particles that reduces both $S_2$ and SOP, the latter effect playing a more dominant role. Finally, we show that the correlation between the local structural parameters ($S_2$ and SOP) and local dynamics increases when the above-mentioned two effects are taken into consideration in the calculation of $S_2$ and SOP.

The rest of the paper is organized as follows: Sec. II contains the simulation details. The analysis at the macroscopic level is presented in Sec. III with excess entropy, $S_{ex}$, in Sec. III A, pair excess entropy, $S_2$, in Sec. III B, the depth of local caging potential, $\beta \Phi_r$, in Sec. III C, and the numerical results in Sec. III D. The analysis at the microscopic level is presented in Sec. IV, with microscopic $S_2$ in Sec. IV A, microscopic $\beta \Phi_r$ in Sec. IV B, and numerical results in Sec. IV C. In Sec. V, we analyze the structure dynamics correlation at the microscopic level. The paper ends with a brief conclusion in Sec. VI. This paper contains seven appendix sections at the end.

## II. SIMULATION DETAILS

In this study, we work with the well-known Kob–Andersen[37] 80:20 binary Lenard-Jones mixture. The shifted and truncated Lennard-Jones interaction potential in the KA model is given by

$$u(r_{\alpha\gamma}) = \begin{cases} u^{(LJ)}(r_{\alpha\gamma}; \sigma_{\alpha\gamma}, \epsilon_{\alpha\gamma}) - u^{(LJ)}(r_{\alpha\gamma}^{(c)}; \sigma_{\alpha\gamma}, \epsilon_{\alpha\gamma}), & r \leq r_{\alpha\gamma}^{(c)}, \\ 0, & r > r_{\alpha\gamma}^{(c)}, \end{cases} \quad (1)$$

where $u^{(LJ)}(r_{\alpha\gamma}; \sigma_{\alpha\gamma}, \epsilon_{\alpha\gamma}) = 4\epsilon_{\alpha\gamma}\left[\left(\frac{\sigma_{\alpha\gamma}}{r_{\alpha\gamma}}\right)^{12} - \left(\frac{\sigma_{\alpha\gamma}}{r_{\alpha\gamma}}\right)^{6}\right]$ and $r_{\alpha\gamma}^{(c)} = 2.5\sigma_{\alpha\gamma}$. Where $\alpha, \gamma \in \{A, B\}$ and $\epsilon_{AA} = 1.0$, $\epsilon_{AB} = 1.5$, $\epsilon_{BB} = 0.5$, $\sigma_{AA} = 1.0$, $\sigma_{AB} = 0.80$, and $\sigma_{BB} = 0.88$. Length, energy, and time scales are measured in units of $\sigma_{AA}$, $\epsilon_{AA}$, and $\sqrt{\frac{\sigma_{AA}^2}{\epsilon_{AA}}}$, respectively. We use a three-dimensional, Lammps-based molecular dynamics simulation.[38] The Nosé–Hoover thermostat is used to simulate NVT molecular dynamics in a cubic box with N = 4000, $\rho = (N/V) = 1.2$, and an integration time step of $\Delta t = 0.005$. The system is equilibrated for a period longer than 100 $\tau_\alpha$, where $\tau_\alpha$ is the system's relaxation time.

The following pinning procedure is applied to create the pinned system: The pinned particles are chosen randomly from an equilibrium configuration of the KA system at the target temperature.[21,39] In this process, we make sure that the ratio of mobile A and mobile B particles in the pin sub-population is the same as the regular KA system (80:20). During random pinning, we maintain the maximum possible distance between two pinned particles. As expected, this maximum distance decreases with an increase in pinning density. The distance between two pinned particles, $r_{PP}$, is greater than 1.9, 1.5, and 1.2 for $c$ = 0.05, 0.10, and 0.15, respectively. The simulations are performed assuming that there is no interaction between two pinned particles ($u_{PP} = 0$). After pinning, we perform an NVT molecular dynamics simulation with an integration time step of $\Delta t = 0.005$. We equilibrate the system for $t = 100$. For this work, we generate three different pinning concentrations "c," i.e., 0.05, 0.10, and 0.15.

In this work, to characterize the dynamics, we consider the self-part of the overlap function, q(t), defined as follows:

$$q(t) = \frac{1}{N}\sum_{i=1}^{N} \omega(|r_i(t) - r_i(0)|), \quad (2)$$

where the function $\omega(x) = 1$ when $0 \le x \le a$ and $\omega(x) = 0$ otherwise. The overlap parameter cutoff ($a$) = 0.3 is taken such that particle positions separated due to small amplitude vibrational motion are treated as the same.[40] We calculate the $\alpha$ relaxation time $\tau_\alpha$ by examining the time where the overlap function decays to $1/e = 0.367$.

## III. ENTROPY AND MEAN FIELD CAGING POTENTIAL AT MACROSCOPIC LEVEL

### A. Macroscopic excess entropy

The excess entropy of a system is the loss of entropy due to the interaction between particles. The excess entropy of pinned systems has been calculated before, and it was also shown that, compared to the unpinned system, the configurational entropy of the system disappears at a higher temperature.[20,22] As discussed in the introduction, this disappearance of the configurational entropy at a temperature where the dynamics continues has been a topic of intense research.[20–22,26,41] The configurational entropy, $S_c = S_{id} + S_{ex} - S_{vib}$, is obtained from the ideal gas entropy, $S_{id}$, excess entropy, $S_{ex}$, and the vibrational entropy, $S_{vib}$, of the system. All these three terms change due to pinning. Here, we first revisit the configurational entropy calculation and find out which terms are primarily responsible for the vanishing of the configurational entropy of the pinned system at a higher temperature.[19,20] As shown in Appendix B, we find that as we increase the pinning concentration, the per particle ideal gas entropy increases. However, the per particle excess entropy and per particle vibrational entropy decrease. The decrease in excess entropy appears to be stronger than vibrational entropy. We make a comparative analysis of the excess entropy of the pinned and unpinned systems to understand what leads to this substantial decrease in the excess entropy.

The excess entropy per particle level is expressed as follows:[30,31]

$$S_{ex}(\beta') = \beta'\langle U \rangle - \int_0^{\beta'} d\beta \langle U \rangle, \quad (3)$$

where $\langle U \rangle$ is the per partial potential energy.

In the case of a regular binary system, the per particle potential energy in terms of the radial distribution function, g(r), can be expressed as follows:[42]

$$\langle U_B \rangle = 2\pi\rho \int_0^\infty \sum_{i=1}^{2}\sum_{j=1}^{2} \frac{N_i}{N}\frac{N_j}{N} u_{ij}(r)g_{ij}(r)r^2 dr$$

$$= 2\pi\rho \int_0^\infty \sum_{i=1}^{2}\sum_{j=1}^{2} \chi_i \chi_j u_{ij}(r)g_{ij}(r)r^2 dr, \quad (4)$$

where $\chi_i = \frac{N_i}{N}$ is the fraction of particles in type $i$. N is the total number of particles in the system.

Note that when we pin particles in a binary system, we actually create a quaternary system of two types of mobile particles and two types of pinned particles. We refer to the first type of mobile particles as species 1, the second type of mobile particles as species 2, the first type of pinned particles as species 3, and the second type of pinned particles as species 4. The potential energy per particle for a regular quaternary system can be expressed as follows:

$$\langle U_Q \rangle = 2\pi\rho \int_0^\infty \sum_{i=1}^{4}\sum_{j=1}^{4} \frac{N_i'}{N}\frac{N_j'}{N} u_{ij}(r)g_{ij}(r)r^2 dr$$

$$= 2\pi\rho \int_0^\infty \sum_{i=1}^{4}\sum_{j=1}^{4} \chi_i' \chi_j' u_{ij}(r)g_{ij}(r)r^2 dr. \quad (5)$$

Now if we assume that a fraction, $c$, of particles are pinned, then $N_1' = (1-c)N_1, N_2' = (1-c)N_2, N_3' = cN_1, N_4' = cN_2, \chi_i' = \frac{N_i'}{N}$. The number of mobile particles can be written as $M = (1-c)N$. In our model system, the pinned particles do not interact with each other;[20] thus, $u_{PP} = u_{33} = u_{34} = u_{43} = u_{44} = 0$. We also know that the interaction between pinned and mobile particles is symmetric; for example, $u_{13} = u_{31}$. These conditions modify the quaternary expression and reduce the first summation in Eq. (5) only over types 1 and 2. Moreover, for a system with pinned particles, the excess entropy, $S_{ex}^M$, is calculated only for the mobile particles, and the total potential energy is divided only between the $M$ mobile particles. This further modifies the quaternary expression [Eq. (5)] and the potential energy at per mobile particle level for the pinned system, which we now also refer to as the modified quaternary system, $\langle U_M \rangle = \frac{N}{M} \times \langle U_Q(u_{PP} = 0) \rangle$ can be written as follows:

$$\langle U_M \rangle = 2\pi\rho \int_0^\infty r^2 dr \sum_{i=1}^2 \frac{N'_i}{M} \left[ \sum_{j=1}^2 \frac{N'_j}{N} u_{ij}(r) g_{ij}(r) \right.$$
$$\left. + 2 \times \sum_{j=3}^4 \frac{N'_j}{N} u_{ij}(r) g_{ij}(r) \right]$$
$$= 2\pi\rho \int_0^\infty r^2 dr \sum_{i=1}^2 \frac{(1-c)N_i}{(1-c)N} \left[ \sum_{j=1}^2 \frac{N'_j}{N} u_{ij}(r) g_{ij}(r) \right.$$
$$\left. + 2 \times \sum_{j=3}^4 \frac{N'_j}{N} u_{ij}(r) g_{ij}(r) \right]$$
$$= 2\pi\rho \int_0^\infty r^2 dr \sum_{i=1}^2 \chi_i \left[ \sum_{j=1}^2 \chi'_j u_{ij}(r) g_{ij}(r) \right.$$
$$\left. + 2 \times \sum_{j=3}^4 \chi'_j u_{ij}(r) g_{ij}(r) \right]. \quad (6)$$

The above expression of the potential energy, when replaced in Eq. (3), provides us with the excess entropy of the mobile particles in the pinned system, $S_{ex}^M(\beta')$. The first and second terms in Eq. (6) describe the potential energy of a mobile particle due to the interaction with other mobile particles and pinned particles, respectively. The expressions of the first and second terms are identical except for the fact that the second term has a factor of 2. This implies that, compared to a mobile particle, a pinned particle has a stronger effect on decreasing the potential energy of a mobile particle. Note that when we pin particles, we can consider that the system is made up of mobile particles in a stationary external field of the pinned particles. Thus, in the calculation of the potential energy, the first summation in Eq. (5) reduces over the first two species, which are the mobile particles. However, the second summation runs over particles that interact with these mobile particles. Thus, it includes both mobile and pinned particles. As shown in Appendix B, in that case, the expression of the potential energy does not have the factor 2 in the second term of $\langle U_M \rangle$ [Eq. (6)] and the excess entropy shows a marginal change. The per particle configurational entropy increases with an increase in pinning density. This is because, with the increase in pinning density, the increase in the ideal gas entropy is greater than the decrease in the vibrational entropy. This result clearly shows that the vanishing of the configurational entropy at higher temperatures is due to the stronger effect of the pinned particles in confining the mobile particles and, thus, decreasing the excess entropy. We will show in Secs. III B and III C that this effect of the pinned particles plays an important role in the two body excess entropy and the mean field caging potential.

### B. Macroscopic pair excess entropy

The excess entropy, $S_{ex}$, can be written in terms of an infinite series via the Kirkwood factorization method,[43,44]

$$S_{ex} = S_2 + S_3 + S_4 \ldots$$
$$= S_2 + \Delta S. \quad (7)$$

While $S_{ex}$ represents the loss of entropy due to total interaction, the pair excess entropy, $S_2$, describes the loss of entropy due to interaction as described by the two-body correlation. $\Delta S$ is the loss of entropy due to many body correlations (beyond pair correlation). The per particle pair excess entropy, which contributes to 80% of the total excess entropy,[29] can be written as follows:[44]

$$\frac{S_2^B}{k_B} = -2\pi\rho \int_0^\infty \sum_{i=1}^2 \sum_{j=1}^2 \chi_i \chi_j \{g_{ij}(r) \ln g_{ij}(r) - (g_{ij}(r) - 1)\} r^2 dr. \quad (8)$$

Pair excess entropy per particle level for the quaternary system is expressed as follows:

$$\frac{S_2^Q}{k_B} = -2\pi\rho \int_0^\infty r^2 dr \sum_{i=1}^4 \sum_{j=1}^4 \chi'_i \chi'_j \{g_{ij}(r) \ln g_{ij}(r) - (g_{ij}(r) - 1)\}. \quad (9)$$

To obtain the pair excess entropy of the pinned system, we make modifications to the pure quaternary system. As done for the excess entropy calculation, the total pair excess entropy is divided only among the mobile particles, and the per particle pair excess entropy of the mobile particles is $S_2^M = \frac{N}{M} * S_2^Q$. Thus, in the first summation, $\chi'_i$ is replaced by $\chi_i$, such as in Eq. (6). The pair excess entropy per particle level of the mobile particles in the pinned system, $S_2^M$, can be written as follows:

$$\frac{S_2^M}{k_B} = -2\pi\rho \int_0^\infty r^2 dr \left[ \sum_{i=1}^2 \chi_i \left[ \sum_{j=1}^2 \chi'_j \{g_{ij}(r) \ln g_{ij}(r) - (g_{ij}(r) - 1)\} \right. \right.$$
$$\left. + 2 \times \sum_{j=3}^4 \chi'_j \{g_{ij}(r) \ln g_{ij}(r) - (g_{ij}(r) - 1)\} \right]$$
$$\left. + \sum_{i=3}^4 \sum_{j=3}^4 \chi_i \chi'_j \{g_{ij}(r) \ln g_{ij}(r) - (g_{ij}(r) - 1)\} \right]. \quad (10)$$

From Eq. (10), we find that, similar to that discussed for excess entropy, when we treat the pinned system as this modified quaternary system, the effect of the pinned particles in determining the entropy of the mobile particles is stronger (by a factor of 2) compared to other mobile particles.

When we pin the particles at their equilibrium position, the structure/rdf of the system is not expected to change. Thus, pinning is believed to keep the equilibrium of the system the same.[24,45,46] If the structure/rdf remains the same, then treating the system as quaternary or binary in the calculation of the two body excess entropy gives us identical results: $S_2^Q = S_2^B$ (see Appendix C). However, note that for the pinned system, the pair excess entropy is not given by $S_2^Q$ [Eq. (9)] but by $S_2^M$ [Eq. (10)]. In the expression of $S_2^M$, even if we assume there is no change in structure due to pinning, the pair excess entropy of the system, $S_2^{M'}$ is different from that of a binary system and can be written as

$$\frac{S_2^{M'}}{k_B} = \frac{N}{M} \left( \frac{S_2^B}{k_B} \right). \quad (11)$$

Since M is always less than N, the above expression suggests that even if the pinning process does not change the structure, the pair excess entropy for mobile particles in the pinned system is lower than that in the unpinned system. This implies that the pinned particles have a stronger confinement effect on the mobile particle. In Sec. III C, we will show that this stronger confining effect of the pinned particles is present not only in entropy but also in other quantities.

### C. Macroscopic mean field caging potential

The time evolution of the density under mean-field approximation can be written in terms of a Smoluchowski equation in an effective mean field caging potential, which is obtained from the Ramakrishnan–Yussouff free energy functional.[32,33,47] Following our earlier studies, the caging potential is calculated by assuming that the cage is static when the particle moves by a distance $\Delta r$.[32] The mean field caging potential is expressed in terms of the static structure factor/radial distribution function of the liquid.[33] In this section, we obtain a pinned system's mean field caging potential. Previous work by some of us showed that the depth of caging potential is coupled to the dynamics.[32,33] Thus, in this study, instead of dealing with the whole potential, we deal with the absolute magnitude of the depth of the caging potential as we view the depth of the caging potential, as an energy barrier. We first start with the binary system, where the average depth of the mean field caging potential can be expressed as follows:[33]

$$\beta \Phi_r^B(\Delta r = 0) = -4\pi\rho \int r^2 dr \sum_{i=1}^{2} \sum_{j=1}^{2} \chi_i \chi_j C_{ij}(r) g_{ij}(r). \quad (12)$$

Here, $r$ is the separation between the tagged particle and its neighbors, $\beta = 1/k_B T$, $k_B = 1$, and $\rho$ is the density. $\Delta r$ is the tagged particle's distance from its equilibrium position. According to the Hypernetted chain approximation, the direct correlation function, $C_{ij}(r)$, can be represented as follows:

$$C_{ij}(r) = -\beta u_{ij}(r) + [g_{ij}(r) - 1] - \ln[g_{ij}(r)]. \quad (13)$$

For a regular quaternary system, the caging potential can be expressed as follows:

$$\beta \Phi_r^Q(\Delta r = 0) = -4\pi\rho \int r^2 dr \sum_{i=1}^{4} \sum_{j=1}^{4} \chi_i' \chi_j' C_{ij}(r) g_{ij}(r). \quad (14)$$

Next, for the calculation of the mean field caging potential for the pinned system, we apply similar conditions as discussed before for the calculation of the excess and pair excess entropies. Under these conditions, the average depth of mean field caging potential of the mobile particles in the pinned system, $\beta \Phi_r^M$ can be written as follows:

$$\beta \Phi_r^M(\Delta r = 0) = -4\pi\rho \int r^2 dr \left[ \sum_{i=1}^{2} \chi_i \left[ \sum_{j=1}^{2} \chi_j' C_{ij}(r) g_{ij}(r) \right. \right.$$
$$\left. + 2 \times \sum_{j=3}^{4} \chi_j' C_{ij}(r) g_{ij}(r) \right]$$
$$\left. + \sum_{i=3}^{4} \sum_{j=3}^{4} \chi_i \chi_j' C_{ij}(r) g_{ij}(r) \right]. \quad (15)$$

Note that, similar to excess and pair excess entropy, the depth of the mean field caging potential of mobile particles in this modified quaternary system is affected more by the pinned particles (a factor of 2) than by other mobile particles. In addition, if the structure does not change due to pinning, the expression of the caging potential for a quaternary and binary system is identical, but that is not the case for the modified quaternary system. The expression for the depth of the mean field caging potential under the assumption that the structure does not change due to pinning can be written as follows:

$$\beta \Phi_r^{M'}(\Delta r = 0) = \frac{N}{M} \beta \Phi_r^B(\Delta r = 0). \quad (16)$$

Since M is always less than N, the above expression suggests that even when we assume that the structure does not change due to pinning, the depth of the caging potential for the pinned system is deeper compared to the unpinned system. This higher confinement effect comes due to the stronger interaction with the pinned particles. Interestingly, a similar effect of the pinned particles has been discussed while studying the nonlinear Langevin equation on a dynamic free energy surface.[35,36] Note that our mean field caging potential is obtained from the functional derivative of the static version of this dynamic free energy.[47,48] Similar to the methodology used here, their study[35,36] on a monoatomic liquid treats the pinned system as a binary system, thus considering the pinned particle as a different species. They also consider the dynamic free energy of only the mobile particles. Under these conditions, they show that the free energy barrier and confinement of the mobile particles increase with pinning density.

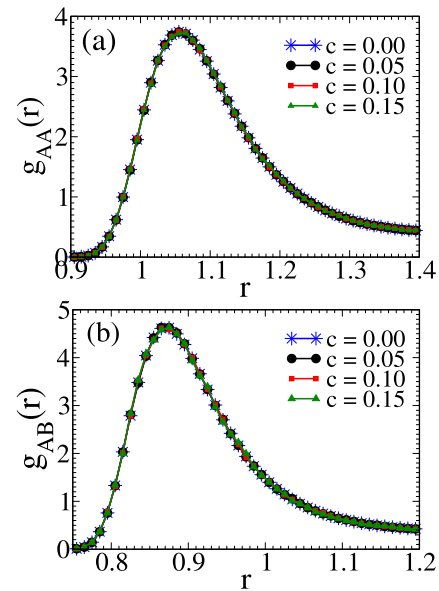

**FIG. 1.** Radial distribution function, $g(r)$, while treating the pinned system as a binary system, at $T = 0.68$. (a) $g_{AA}$ as a function of $r$. (b) $g_{AB}$ as a function of $r$. Here, A and B are the bigger and smaller sizes of particles, respectively.

### D. Numerical results for the macroscopic pair excess entropy and mean field caging potential

Note that the two body excess entropy and the mean field caging potential are both functions of the radial distribution function (rdf) given by

$$g_{ij}(r) = \frac{V}{N_i N_j} \left\langle \sum_{\alpha=1}^{N_i} \sum_{\beta=1, \beta \neq \alpha}^{N_j} \delta(r - r_\alpha + r_\beta) \right\rangle, \quad (17)$$

where V is the system's volume and $N_i$ and $N_j$ are the number of particles of the $i$ and $j$ types, respectively. $r_\alpha$ and $r_\beta$ are the $\alpha$th and $\beta$th particles' positions in the system, respectively.

In Fig. 1, we plot the partial rdfs of the system where we do not differentiate between the pinned and unpinned particles, and we find that, as expected, the rdf remains the same as the unpinned regular KA model (c = 0).

In the rest of the article, when we refer to the unpinned binary KA system, following the usual norm, we refer to the particles as A and B types. However, as discussed in the previous sections, when we pin particles in a binary system, we actually create a quaternary system. We refer to mobile A type of particles as 1, mobile B type of particles as 2, pinned A type of particles as 3, and pinned B type of particles as 4.

We next plot some representative partial rdfs, assuming the system to be quaternary at different pinning concentrations (Fig. 2). We find that with increased pinning density, the partial rdfs start deviating from the c = 0 system. With an increase in "c," there is a drop in the peak value of the rdfs between two mobile particles ($g_{11}$). On the other hand, the first peak height of the partial rdfs between mobile and pinned particles ($g_{13}$) grows with "c." Details are shown in Appendix D. To ensure that this is not an artifact of choosing the pinned particles as a different species, in the c = 0 system, we randomly choose 15% of the particles and treat them as a different species. In Fig. 2, we show that in that case, $g_{11} = g_{13} = g_{AA}$. A similar result is also observed for other partial rdfs (not shown here). This clearly shows that when we pin a certain fraction of particles, contrary to common belief, there is a structural change.

We observe that this structural change happens quickly, immediately after the pinning process. We calculate $g(r)$, averaged from $t = 0$–100 and $t = 101$–200, where the pinning is performed at $t = 0$. We find that both rdfs overlap (Appendix D, Fig. 14). In Appendix D, Fig. 15, we also show that $\chi'_1 g_{11} + \chi'_3 g_{13}$ is the same as $\chi_A g_{AA}$ and $\chi'_2 g_{12} + \chi'_4 g_{14}$ is the same as $\chi_B g_{BB}$. This is precisely why we do not see a change in structure when the pinned particles are not treated as a different species (Fig. 1). Note that this change in the partial rdfs is independent of the integration timestep and system size (Appendix D, Fig. 17).

In the earlier studies of the system with pinned particles, the dynamics and entropy were calculated only for the subsystem of the mobile particles;[20–22,26,27,41] however, for the structure and structural order parameters, the whole system was considered, and it was suggested that there was no change in structure due to pinning.[24] Our present study shows that, even when we consider the structure of the whole system, we have to be careful and treat the pinned particles as a different species, and in doing so, we observe a change in the structure of the system. This change is significant at higher pinning densities. We should also emphasize the fact that the structure depends on the pinning protocol. When the pinning is done without any restriction on the distance between the pinned particles, there is no change in structure, even when the pinned particles are treated as a different species (not shown here).

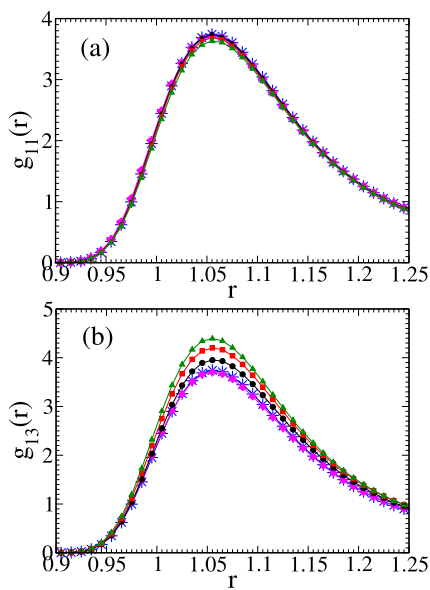

**FIG. 2.** Comparison between radial distribution functions, g(r)s, by randomly picking 15% of the particles in the KA system and treating them as different species (magenta diamond) and pinning 15% of the particles' position during the simulation and treating the pinned particles as different species (c = 0.05, black circle; c = 0.10, red square; and c = 0.15, green triangle). We also plot the g(r) for a regular KA (c = 0) system for comparison (blue, star). (a) Radial distribution function between mobile A and mobile A [$g_{11}(r)$]. (b) Radial distribution function between mobile A and pinned A [$g_{13}(r)$].

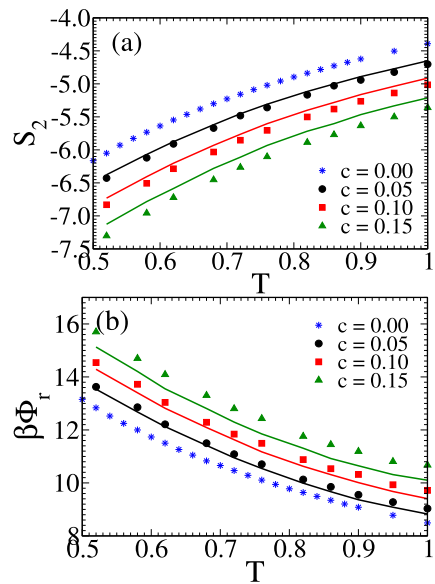

**FIG. 3.** (a) Macroscopic $S_2$ as a function of T. (b) Macroscopic $\beta\Phi_r$ as a function of T. A solid line represents $S_2^{M'}$ (Eq. 11) or $\beta\Phi_r^{M'}$ [Eq. (16)] and symbol represents $S_2^M$ (Eq. 10) or $\beta\Phi_r^M$ [Eq. (15)].

In Fig. 3, we plot the pair excess entropy, $S_2^{M'}$ [Eq. (11)] of the pinned systems, where we assume that the structure does not change due to pinning. We find that, even when we consider that the structure does not change, the pair excess entropy of the pinned system differs from that of the binary system and decreases with increasing pinning density. We also plot $S_2^M$ [Fig. 3(a)], where we consider that the structure changes due to pinning. We find that the entropy further decreases, and similar to the structure, this effect is significant at higher pinning densities. As discussed earlier, the change in structure is dependent on the pinning protocol. However, our analysis clearly shows that the change in the analytical expression of the pair excess entropy is independent of the pinning protocol and depends on the fraction of pinned particles. Thus, our study suggests that, even if the structure changes in a perturbative manner, the change in the structural parameters is non-perturbative. The plot of the mean field caging potential $\beta\Phi_r^{M'}$ [Eq. (16)] [Fig. 3(b)] shows a similar effect. The caging potential depth increases with pinning, even if the change in the structure due to pinning is ignored. There is a further increase in depth when the change in structure is taken into account.

Thus, we find that both the pair excess entropy and the mean field caging potential for the pinned system differ from that of the unpinned system, and this difference comes due to two factors. First, compared to a mobile particle, a pinned particle has a stronger effect on the mobile particle, leading to a decrease in entropy and an increase in caging potential. Second, due to pinning, the structure of the liquid changes, and this further decreases the entropy and increases the mean field caging potential. As shown in Fig. 3, the first effect is stronger, non-perturbative in nature, independent of the pinning protocol, and plays a dominant role.

In Appendix C, we show that the well-known crossover[49] between the excess entropy and the pair excess entropy takes place at a physically meaningful temperature only when we take into consideration these two effects in the calculation of the entropy.

## IV. PAIR EXCESS ENTROPY AND MEAN FIELD CAGING POTENTIAL AT THE MICROSCOPIC LEVEL

In Sec. III D, we developed the protocol for calculating the caging potential and pair excess entropy at the macroscopic level for the pinned system. However, our primary goal is to understand how these two order parameters can describe the dynamics at the local level. We clearly demonstrate that the pinned system should be treated as a modified quaternary system. In this section, we make a comparative analysis of these two structural quantities when the pinned system is treated as a binary system and a modified quaternary system. First, we start with the microscopic expressions, which are obtained from the macroscopic expressions. The bigger "A" particles, which are larger in number, are the ones for which all microscopic calculations are performed. This is done to make sure that there is no size inhomogeneity, which we know also affects the dynamics.[50]

### A. Microscopic pair excess entropy

The calculation of the pair excess entropy at the macroscopic level ($S_2$) is given in Sec. III.

In the binary system, the pair excess entropy of each mobile "A" particle, which is type "1" in our notation, can be expressed by removing the first summation in Eq. (8),

$$\frac{S_2^B(A)}{k_B} = -2\pi\rho \int_0^\infty r^2 dr \sum_{j=1}^2 \chi_j \{g_{1j}(r) \ln g_{1j}(r) - (g_{1j}(r) - 1)\}. \quad (18)$$

Similarly, in the modified quaternary system, the pair excess entropy of each mobile "A" particle (type 1) can be expressed by removing the first summation in Eq. (10),

$$\frac{S_2^M(A)}{k_B} = -2\pi\rho \int_0^\infty r^2 dr \left[ \sum_{j=1}^2 \chi_j' \{g_{1j}(r) \ln g_{1j}(r) - (g_{1j}(r) - 1)\} \right.$$
$$\left. + 2 \times \sum_{j=3}^4 \chi_j' \{g_{1j}(r) \ln g_{1j}(r) - (g_{1j}(r) - 1)\} \right]. \quad (19)$$

In addition, note that, as we are not calculating the entropy of the pinned particles, we drop the last term in Eq. (10). The differences between binary and modified quaternary are the following: In the binary expression, when treating the neighbors, we do not differentiate between the mobile and pinned particles; however, in the quaternary expression, we do. Thus, in the binary expression, the effect of the mobile neighbors on the tagged particle is the same as that of the pinned neighbors. However, in the quaternary expression, the effect of the pinned neighbors on the tagged particle is twice that of the mobile neighbors. As shown in the macroscopic calculation (Fig. 3), it is this second effect that plays a dominant role in differentiating between the binary and the modified quaternary values of the entropy.

### B. Microscopic mean field caging potential

The macroscopic calculation of the depth of the caging potential ($\beta\Phi_r$), the inverse of which we refer to as the structural order parameter (SOP), is given in Sec. III C. At the microscopic level for a binary system, the caging potential of a mobile "A" type of particle can be written by removing the first summation in Eq. (12),

$$\beta\Phi_r^B(A, \Delta r = 0) = -4\pi\rho \int r^2 dr \sum_{j=1}^2 \chi_j C_{1j}(r) g_{1j}(r). \quad (20)$$

The mean field caging potential for a mobile "A" type of particle in a modified quaternary system can be written by removing the first summation in Eq. (15),

$$\beta\Phi_r^M(A, \Delta r = 0) = -4\pi\rho \int r^2 dr \left[ \sum_{j=1}^2 \chi_j' C_{1j}(r) g_{1j}(r) \right.$$
$$\left. + 2 \times \sum_{j=3}^4 \chi_j' C_{1j}(r) g_{1j}(r) \right]. \quad (21)$$

Note that, similar to that discussed for pair excess entropy, since we are not calculating the caging potential of the pinned particles in the above expression, we drop the last term in Eq. (15). Thus, in the modified quaternary expression, compared to the mobile neighbors, the pinned neighbors have a stronger effect on confining the tagged particle.

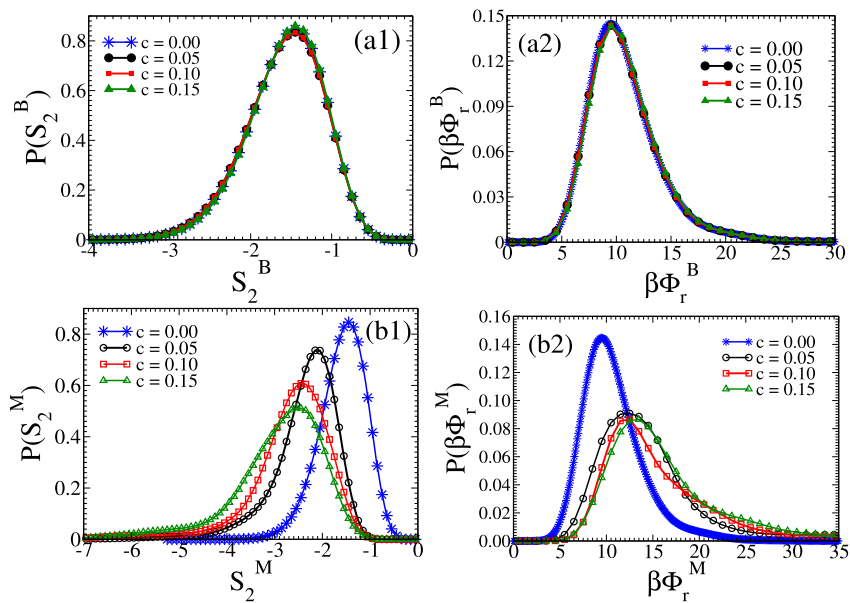

**FIG. 4.** Distribution of pair excess entropy ($S_2$) and depth of mean field caging potential ($\beta\Phi_r$) in different pinned systems at T = 0.68. (a1) Distribution of $S_2$ treating the pinned system as binary. (a2) Distribution of $\beta\Phi_r$ treating the pinned system as binary. (b1) Distribution of $S_2$ treating the pinned system as modified quaternary. (b2) Distribution of $\beta\Phi_r$ treating the pinned system as modified quaternary. The distribution remains the same for the binary treatment, while for the modified quaternary treatment, the caging potential increases with increasing c, and the pair excess entropy decreases with increasing c.

### C. Numerical results for the microscopic pair excess entropy and mean field caging potential

To perform the microscopic investigation, we determine $\beta\Phi_r(\Delta r = 0)$ and $S_2$ for every snapshot at the single particle level that requires the partial rdfs to be calculated at a single particle level. In this calculation, the sum of the Gaussian can be used to express the single particle partial rdf in a single frame, and it is calculated as follows:[51]

$$g_{ij}^{\alpha}(r) = \frac{1}{4\pi\rho r^2}\sum_{\beta}\frac{1}{\sqrt{2\pi\delta^2}}\exp^{-\frac{(r-r_{\alpha\beta})^2}{2\delta^2}}, \quad (22)$$

where "$\alpha$" is the particle index and $\rho$ is the density. The Gaussian distribution's variance ($\delta$) is employed to transform the discontinuous function into a continuous form. We use $\delta = 0.09\sigma_{AA}$ for this work. Single particle rdf is used to derive the direct correlation function at the single particle level from Eq. (13).

We can determine caging potential [Eqs. (12), (14) and (15)] by combining the direct correlation function [Eq. (13)] and particle level rdf [Eq. (22)]. This leads to a term that is a product of the interaction potential and the rdf. As shown in an earlier work,[33] at distances shorter than the average rdf, the particle level rdf generated by the Gaussian approximation has finite values. At small "r," due to this finite value of the rdf, its product with the interaction potential, which diverges at small "r," leads to a large unphysical contribution from this range. To get around this problem, we use an approximate expression of the direct correlation function, $C^{approx}(r) = [g_{ij}(r) - 1]$, where we assume that the interaction potential is equal to the potential of mean force $-\beta u_{ij}(r) = \ln(g_{ij}(r))$. It has also been shown earlier that using $C_{ij}^{approx}(r)$ marginally improves the theoretical prediction of structure-dynamics correlation.[34,50] In the rest of the microscopic calculation, we will use $C_{ij}^{approx}(r)$ in place of $C_{ij}(r)$.

In Fig. 4, we plot the distribution of pairs' excess entropy and local caging potential by describing the pinned system as binary. Note that for all the cases, the quantities are calculated only for mobile "A" particles. We find that the distribution remains similar to the unpinned system (KA model at c = 0). This is because the expressions are identical for pinned and unpinned systems, and even for c = 0.15, there are enough mobile "A" particles to give the correct statistics. However, when we calculate the quantities assuming the pinned system is a modified quaternary system [Eqs. (15) and (10)], we observe that as "c" increases, the depth of the caging potential increases and the pair excess entropy decreases. The distribution of $\beta\Phi_r^M$ and $S_2^M$ are shown in Fig. 4. This analysis clearly shows that the entropy and the SOP (inverse depth of the caging potential) are higher when the system is treated as binary compared to when it is treated as a modified quaternary. In the next section, we will show that the correlation between the dynamics and structural quantities differs when we treat the pinned system as binary or modified quaternary.

## V. CORRELATION BETWEEN STRUCTURE AND DYNAMICS AT MICROSCOPIC LEVEL

In the following section, we study the correlation between two structural order parameters, namely, the $S_2$ and SOP, with the

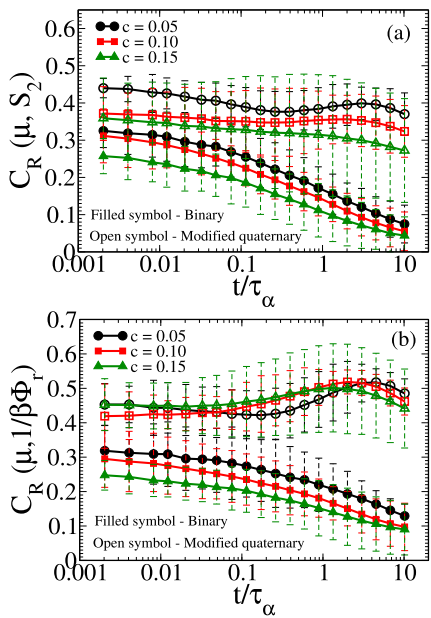

**FIG. 5.** Spearman rank correlation, $C_R$, between different parameters while treating the pinned system as binary (filled symbol) and modified quaternary (open symbol). (a) Spearman rank correlation ($C_R$) between mobility ($\mu$) and pair excess entropy ($S_2$). (b) Spearman rank correlation ($C_R$) between mobility ($\mu$) and inverse depth of caging potential ($1/\beta\Phi_r$). Working temperatures for c = 0.05, 0.10, and 0.15 are 0.52, 0.60, and 0.68, respectively. Note that T is chosen such that all the pinned systems have approximately the same $\tau_\alpha \approx 10^3$.

We calculate the Spearman rank correlation, $C_R(X, Y) = 1 - \frac{6 \sum d_i^2}{m(m^2-1)}$ [where $d_i^2 = R(X_i) - R(Y_i)$ is the difference between the ranks, $R(X_i)$ and $R(Y_i)$, of the raw data $X_i$ and $Y_i$, respectively; and m denotes the number of data], between the mobility, $\mu$, and the pair excess entropy $[C_R(\mu, S_2)]$, and between the mobility, $\mu$, and the SOP $C_R(\mu, 1/\beta\Phi_r)$. In Figs. 5(a) and 5(b), we plot $C_R(\mu, S_2)$ and $C_R(\mu, 1/\beta\Phi_r)$, respectively, for the pinned systems as a function of scaled time. We observe that when considering the system as a binary system, the correlations, $C_R(\mu, S_2^B)$ and $C_R(\mu, 1/\beta\Phi_r^B)$ decrease as the pinning concentration increases (Fig. 5). This observation is concurrent with the findings of Williams et al.[24] However, when the system is treated as a modified quaternary system, we observe an increase in $C_R(\mu, S_2^M)$ and $C_R(\mu, 1/\beta\Phi_r^M)$ compared to when the system is treated as binary. This suggests that treating the system as binary does not capture the full complexity of the structure-dynamics relationship. In the modified quaternary treatment of the system, the pinning decreases the pair excess entropy and the SOP, which is commensurate with the slowing down of the dynamics.

Between the SOP and the pair excess entropy, we find that the SOP can predict the dynamics better and $C_R(\mu, 1/\beta\Phi_r^M) > C_R(\mu, S_2^M)$. This is similar to that observed in an earlier study where, for attractive systems compared to entropy, the SOP is a better predictor of the dynamics.[34] In addition, note that for all values of "c," the peak height of the $C_R(\mu, 1/\beta\Phi_r^M)$ almost remains constant, whereas the peak height of $C_R(\mu, S_2^M)$ drops with an increase in "c." Thus, the difference between $C_R(\mu, 1/\beta\Phi_r^M)$ and $C_R(\mu, S_2^M)$ increases with "c." This drop in the value of $C_R(\mu, S_2^M)$ with an increase in "c" may be connected to the breakdown of the AG relationship at the macroscopic level. However, we cannot calculate the configurational entropy at the microscopic level, but we do find in Fig. 4 that the shift in the distribution of the pair excess entropy with pinning density is stronger than the shift in the distribution of SOP.

We also find that with increasing pinning concentration, the peak height of $C_R(\mu, 1/\beta\Phi_r^B)$ moves to smaller values of $t/\tau_\alpha$. A similar observation was made while comparing the more fragile Lennard-Jones (LJ) and the less fragile Weeks–Chandler–Anderson (WCA) models.[34] Note that in the case of pinned systems, the fragility decreases with increasing "c."[21] Thus, it appears that for more fragile systems, the correlation between structure and dynamics continues for longer periods of time. However, at this point, this dynamics using different techniques. To make a comparative analysis, while calculating the structural quantities, we treat the pinned system both as binary and modified quaternary systems.

### A. Correlation between structure and dynamics using isoconfiguration runs

In this section, we study the correlation between structure and dynamics using isoconfiguration runs (IC). IC is a powerful technique developed by Harrowell et al.[52–55] to examine the role the structure plays in the dynamics (details are given in Appendix E).

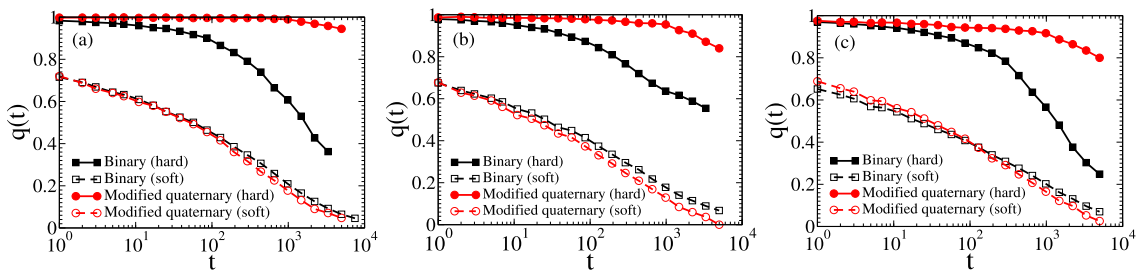

**FIG. 6.** Comparison of the dynamics of a few hardest (filled symbol) and a few softest (open symbol) particles at different pinning concentrations while treating the pinned system as binary (black), and modified quaternary (red). (a) c = 0.05 (at T = 0.52), (b) c = 0.10 (at T = 0.60), and (c) c = 0.15 (at T = 0.68). Note that T is chosen such that all pinned systems have approximately the same $\tau_\alpha \approx 10^3$.

is only a conjecture, and to make more concrete statements, further investigations are needed, which is beyond the scope of the present study.

### B. Analysis of dynamics of particles belonging to the softest and hardest regions

Since we show that the inverse of the mean field caging potential, SOP, is a better predictor of the dynamics, in the next two subsections, we will present the study of the structure-dynamics correlation using only the SOP. At short timescales, we expect to observe a significant difference in the dynamics of the hardest (in a deep cage) and softest (in a shallow cage) particles. The hardest particles, which are less likely to escape their local cages, will exhibit slower dynamics. On the other hand, the softest particles, with a higher probability of moving, will display faster dynamics. However, over a longer period of time, as the cage evolves, the separation in dynamics between the hardest and softest particles diminishes.[32,33,50] We average over a few (∼2 to 3) hardest and softest particles and compare their dynamics via the overlap function [Eq. (2)]. Note that the identity of the soft and hard particles depends on how the SOP is calculated, i.e., assuming the system to be binary or modified quaternary.

The dynamics of the hardest and softest particles for different concentrations of pinning are shown in Fig. 6. When we calculate the SOP treating the system as a modified quaternary system, the difference in dynamics between the hard and soft particles is wider compared to the case where the system is treated as binary (Fig. 6). Note that the difference is greater for the hard particles. This is because our analysis reveals that the identity of the softest particles does not change when we treat the system as binary or modified quaternary. However, the identity of the hardest particles completely changes because, in the binary treatment, we neglect the stronger interaction between the pinned and the mobile particles, which is present in the modified quaternary treatment. Due to this effect in the modified quaternary treatment, the hardest particles are the ones that have more pinned particles as their neighbors. This can be seen in Fig. 7, where we plot the probability of $n_P$, the number of pinned neighbors in the first shell of the hard particles, for c = 0.5 and c = 0.15. We find that, compared to the case where we treat the system as a binary system, when we treat the system as a modified quaternary system, the probability moves to higher values of $n_p$. We also find that the hardest particles, as identified by the modified quaternary treatment, are slower than those identified by the binary treatment (Fig. 6). This is precisely the reason that in the modified quaternary treatment, the system shows a higher value of $C_R(\mu, 1/\beta\Phi_r^M)$ compared to the binary treatment of the system.

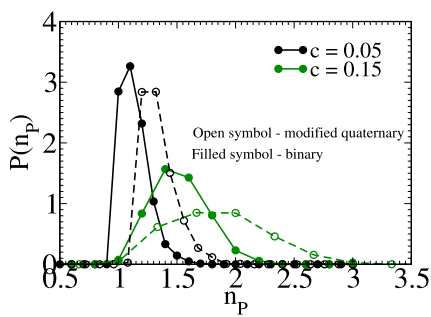

**FIG. 7.** Probability distribution of $n_P$, the number of pinned neighbors in the first shell of the ten hardest particles.

In Appendix G, we illustrate the correlation between structure and dynamics and the prediction of onset temperature. We note that when we treat the system as binary, the onset temperature remains the same. However, when the pinned system is treated as a modified quaternary system, we observe an increase in the onset temperature with pinning. A similar effect is also observed when the onset temperature is calculated using the inherent energy method.[56] Thus, all the above analysis suggests that when we treat the pinned system as a modified quaternary system, the structure-dynamics correlation is higher.

## VI. CONCLUSION

As discussed in the introduction, earlier studies on the pinned system have shown that both at macroscopic and microscopic levels, the correlation between dynamics and entropy breaks down. However, the nature of the breakdown at the microscopic and macroscopic levels is not similar but just the opposite. At the macroscopic level, with pinning, the configurational entropy disappears, whereas the dynamics continues.[20,22,26] At the microscopic level, the pair excess entropy remains high and the same as in the unpinned system, whereas the dynamics slow down with an increase in pinning density.[24] This is possible only when the macroscopic configurational entropy and the microscopic pair excess entropy are uncorrelated. However, it is well known that for unpinned systems, the pair excess entropy contributes to about 80% of the excess entropy, which in turn contributes to the configurational entropy.[29] Thus, to understand the different results at the macroscopic and microscopic levels, we revisit the excess entropy calculation of the pinned system.

We show that when we pin particles in a binary system, we should treat this pinned system as a quaternary system under the assumption that there is no interaction between pinned particles (an assumption we use while simulating the system) and the potential energy is distributed only among the mobile particles. The excess entropy of this modified quaternary system predicts that the effect of a pinned particle in stabilizing a mobile particle by decreasing the potential energy is a factor of two more than the effect of another mobile particle. We show that this effect leads to the well-documented vanishing of configurational entropy at higher temperatures[19] and the breakdown of the Adam–Gibbs relationship in a pinned system.[20,22]

We follow the same logic to calculate the pair excess entropy and the mean field caging potential at macroscopic and microscopic levels. We first show that the expressions of $S_2$ and SOP (inverse depth of the mean field caging potential) differ when the system is treated as binary and modified quaternary. In the binary treatment, the effect of a pinned particle on the mobile particle is identical to that of another mobile particle. However, in modified quaternary treatment, similar to that observed in the calculation of the excess entropy, the pinned particles have a stronger effect on the mobile particles than other mobile particles. We next show that contrary to the common belief that if pinned at the equilibrium position, the properties of the system do not change, pinning changes the structure of the liquid, which can be observed only when we treat the pinned particles as a different species. We then show that when we treat the system as a modified quaternary system, the entropy and the SOP are much lower than those obtained by treating the system as a binary system. The analysis reveals that, more than the change in structure, the stronger effect of the pinned particles on the mobile particles plays a dominant role in confining the mobile particles by decreasing the entropy and the SOP. Interestingly, a similar confinement effect of the pinned particles was discussed in an earlier study of a monotonic system, where it was shown that the free energy barrier of the mobile particles increases with pinning density.[35,36] Note that, similar to the present study, in these earlier studies,[35,36] the pinned particles were treated as a different species.

We further study the correlation between structure and dynamics using different techniques. In all cases, we show that, compared to the case where the pinned system is treated as a binary system, there is an increased correlation between structural order parameters and the dynamics when the pinned system is treated as a modified quaternary system. This is because, unlike in the binary case, in the modified quaternary case, the pinned particles affect not only the dynamics but also the structural properties. We also show that, compared to the entropy, the SOP can predict the dynamics better. The correlation between fast particles and the SOP can only predict the correct onset temperature when the SOP is calculated, assuming the pinned system is a modified quaternary system.

Note that many glass forming systems are not binary but polydisperse in nature,[57–64] and as shown in our earlier study, the way any structural parameter is defined in these systems is nontrivial.[50] Depending on the degree of polydispersity, the polydisperse system should be treated as a $M_0$ number of species system.[65–67] Our present study suggests that when we pin particles in a polydisperse system and choose the pinned particle size distribution in a similar way as the total number of particles, the system should be treated as a $2 \times M_0$ species. To get the correct structure dynamics correlation in these systems, we should again take into consideration the stronger confining effect of the pinned particles on the mobile ones.

In summary, our study reveals a few important observations. The pinning affects not only the dynamics but also the structural and thermodynamical parameters both at macroscopic and microscopic levels. The effect of pinning on the structural parameters is observed in their analytical expressions when we treat the pinned particles as different species. The expressions suggest that, compared to another mobile particle, a pinned particle has a stronger effect on a mobile particle. We also find that, with our pinning protocol, the structure of the liquid changes. These two contributions lead to a substantial change in the values of the structural order parameters. Thus, the study reveals that the pinning process may have a perturbative effect on the structure of the liquid but has a non-perturbative effect on the structural order parameters, which we believe is a non-intuitive and important result.

## ACKNOWLEDGMENTS

P.P. acknowledges CSIR for the research fellowships. S.M.B. acknowledges SERB (Grant No. SPF/2021/000112) for the funding. The authors would like to thank Chandan Dasgupta, Smarajit Karmakar, Ujjwal Kumar Nandi, Mohit Sharma, and Manoj Kumar Nandi for the discussions.

## AUTHOR DECLARATIONS

### Conflict of Interest

The authors have no conflicts to disclose.

### Author Contributions

**Palak Patel**: Data curation (equal); Formal analysis (equal); Methodology (equal); Software (lead); Validation (equal); Visualization (equal); Writing – original draft (equal); Writing – review & editing (equal). **Sarika Maitra Bhattacharyya**: Conceptualization (lead); Formal analysis (equal); Funding acquisition (lead); Methodology (equal); Project administration (lead); Resources (lead); Supervision (lead); Validation (lead); Visualization (equal); Writing – original draft (lead); Writing – review & editing (lead).

## DATA AVAILABILITY

The data that support the findings of this study are available from the corresponding author upon reasonable request.

## APPENDIX A: ONSET TEMPERATURES OF THE PINNED SYSTEMS FROM INHERENT STRUCTURE ENERGY

To estimate the temperature range of the system, we first obtain the onset temperature of the supercooled. In Fig. 8, we plot the inherent structure energy, $e_{IS}$, as a function of T to calculate the onset temperature ($T_{onset}$) from the inherent structure (IS).[56] $T_{onset}$ at different pinning concentrations is given in Table I. The IS is obtained using the FIRE algorithm.[68] From this analysis, we observe that the onset temperature increases with increasing pinning concentration, "c"[19] (see Table I).

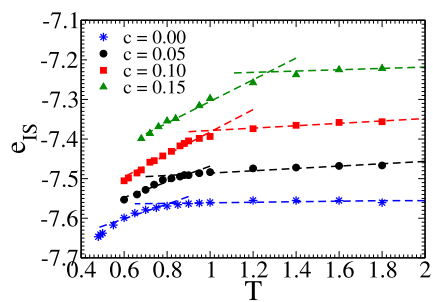

**FIG. 8.** Inherent structure energy, $e_{IS}$, as a function of temperature (T). The onset temperature is the temperature where $e_{IS}$ starts to drop from its high-temperature value. Onset temperature increases as c increases.

**TABLE I.** Onset temperature, $T_{onset}$, at different pinning concentrations, "c."

| c | $T_{onset}$ |
| --- | --- |
| 0.00 | 0.80 |
| 0.05 | 0.89 |
| 0.10 | 1.01 |
| 0.15 | 1.27 |

## APPENDIX B: VARIOUS FORMS OF ENTROPY IN PINNED SYSTEMS

The various forms of entropy in pinned systems are discussed here.

- **Ideal gas entropy:** The ideal gas entropy in pinned systems only comes from the moving particle. The pinned system's ideal gas entropy is calculated as follows:[20]

$$MS_{id} = \frac{3M}{2} - 3M \ln \Lambda + M\left(1 - \ln \frac{M}{V}\right) - \sum_{i=1}^{2} N'_i \ln \frac{N'_i}{M}, \quad (B1)$$

where $M$ is the total number of particles that are moving and $\Lambda = \sqrt{\frac{\beta h^2}{2\pi m}}$ is the de Broglie thermal wavelength, and h is the Planck constant. We plot the per particle ideal gas entropy of pinned systems at various pinning concentrations in Fig. 9. As the pinning increases, we see an increase in ideal gas entropy. The decrease in the density (as $M < N$) and the increase in the mixing entropy contribute to the increase in the per particle ideal gas entropy.

- **Vibrational entropy:** We consider a weakly vibrating system around an inherent structure (IS). If we indicate by $r_i$ the displacement of the $i$th particle from its point in the IS, then the potential energy can be approximated well by the following formula:[20]

$$U \approx U_{IS}(S) + \frac{1}{2} \sum_{i,j}^{M} \frac{\delta^2 U}{\delta r_i \delta r_j} \delta r_i \delta r_j. \quad (B2)$$

It is important to realize that only the derivative of the potential energy with respect to the coordinates of unpinned particles should be taken into account, not including the ones of pinned particles. (However, of course, U will depend on the positions of the pinned and unpinned particles.) Thus, the size of the Hessian matrix is $3M \times 3M$. Introducing the eigenvalues $\lambda_1, \lambda_2 \ldots \lambda_{3M}$ of the Hessian, the harmonic vibrational entropy of the given inherent structure with a given pinned particle configuration can be written as follows:[20]

$$MS_{vib} = 3M(1 - \ln \Lambda) + \frac{1}{2}\sum_{i=1}^{3M} \ln \frac{2\pi}{\beta m \omega_i^2}. \quad (B3)$$

We plot the vibrational entropy of pinned systems at various pinning concentrations in Fig. 9. As the pinning increases, we see a drop in vibrational entropy.

- **Excess entropy:** We employ the thermodynamic integration approach to determine entropy from simulations. At the target temperature $\beta'$, the entropy of the system with the pinned particles S can be written as follows:[20,22,31]

$$S_{ex}^M(\beta') = \beta'\langle U_M \rangle - \int_0^{\beta'} d\beta \langle U_M \rangle, \quad (B4)$$

where $\langle U_M \rangle$ is a thermal average of the potential energy. Details of the excess entropy calculation are discussed in Sec. III A. We plot the excess entropy of pinned systems at various pinning concentrations in Fig. 9. As the pinning increases, we see a drop in excess entropy. We also plot the excess entropy for the pinned system, where we consider the pinned particles to be analogous to a stationary external field. In that case, in the expression of the potential energy, the first summation will be only over the mobile particles, and we will not have the prefactor 2 in the second term of $\langle U_M \rangle$ [Eq. (6)]. The potential energy $\langle U_M^I \rangle$ of the system where the pinned particles are assumed to be analogous to a stationary external potential is expressed as follows:

$$\langle U_M^I \rangle = 2\pi\rho \int_0^\infty r^2 dr \sum_{i=1}^{2} \chi_i \left[ \sum_{j=1}^{2} \chi_j' u_{ij}(r) g_{ij}(r) + \sum_{j=3}^{4} \chi_j' u_{ij}(r) g_{ij}(r) \right]. \quad (B5)$$

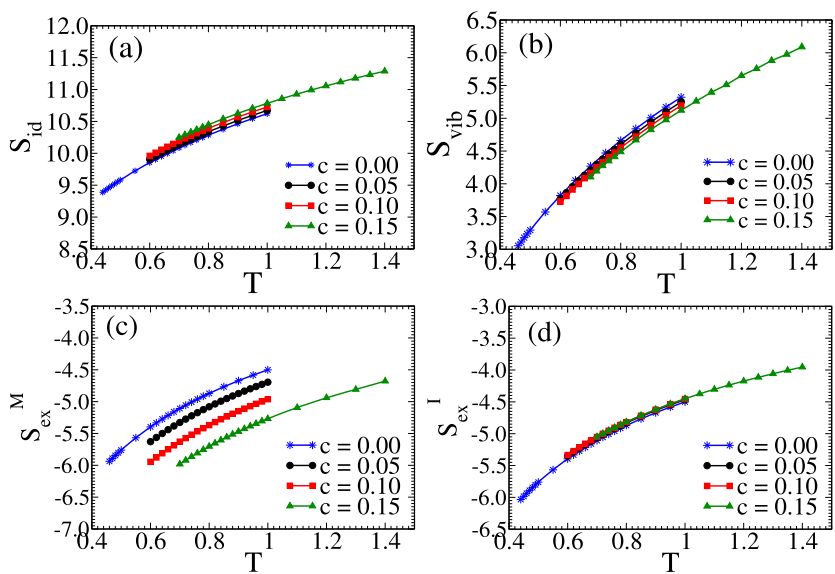

**FIG. 9.** Various forms of entropy as a function of temperature, T. (a) Ideal gas entropy, $S_{id}$. (b) Vibrational entropy $S_{vib}$. (c) Excess entropy, $S_{ex}^M$ [Eq. (B4)]. (d) Excess entropy, $S_{ex}^I$ [Eq. (B6)].

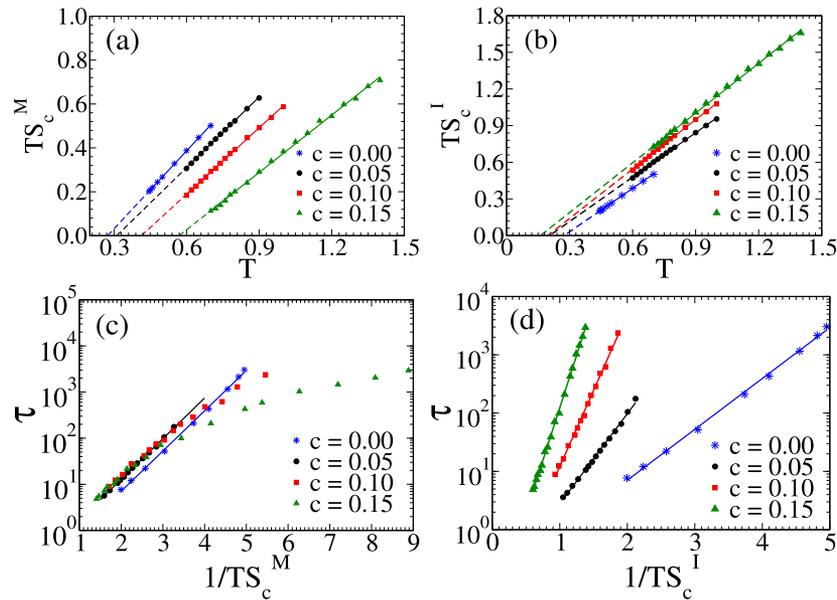

**FIG. 10.** (a) $TS_c^M = T(S_{id} + S_{ex}^M - S_{vib})$, as a function of T. (b) $TS_c^I = T(S_{id} + S_{ex}^I - S_{vib})$, as a function of T. (c) $\tau$ vs $1/TS_c^M$. The solid lines show the Adam–Gibbs fitting. (d) $\tau$ vs $1/TS_c^I$. The solid lines show the Adam–Gibbs fitting.

In this case, the excess entropy can be written as follows:

$$S_{ex}^I = \beta' \langle U_M^I \rangle - \int_0^{\beta'} d\beta \langle U_M^I \rangle. \quad (B6)$$

We find that the excess entropy, $S_{ex}^I$, does not decrease with pinning. Rather, it shows a marginal increase. This analysis clearly shows that the decrease in the excess entropy with pinning is due to the higher potential energy contribution of the pinned particles, which leads to a stronger confinement of the mobile particles.

In Fig. 10, we plot the configurational entropy, $S_c^M = S_{id} + S_{ex}^M - S_{vib}$ and $S_c^I = S_{id} + S_{ex}^I - S_{vib}$ at different pinning concentrations. We observe that the Kauzmann temperature, $T_K$, where the extrapolated entropy vanishes, increases when excess entropy is calculated using the modified quaternary expression of the potential energy, $\langle U_M \rangle$, and the Adam–Gibbs relation between the dynamics and entropy is not valid. However, when we treat the pinned particles as an external stationary field, i.e., use $\langle U_M^I \rangle$ in the calculation of the excess entropy, $T_K$ decreases with increasing pinning, and the Adam–Gibbs relation between the dynamics and entropy is valid. This analysis clearly shows that in the pinned system, the vanishing of the entropy at higher temperatures is due to the stronger confinement effect of the pinned particles on the mobile particles.

### APPENDIX C: PAIR EXCESS ENTROPY

In Sec. III B, we show that the pair excess entropy can have different expressions when the system is treated as binary, quaternary, and modified quaternary. We also show that the rdf is different when the system is treated as binary and quaternary (Sec. III D).

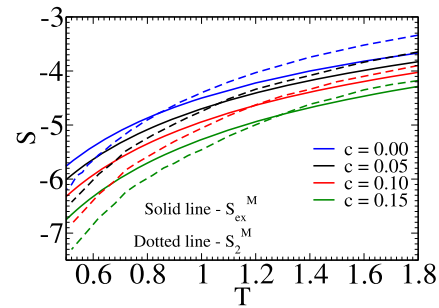

**FIG. 11.** Crossing between pair excess entropy $S_2^M$ [Eq. (10)] and excess entropy $S_{ex}^M$ [Eq. (B4)].

If the structure (rdf) does not change, then treating the system as quaternary or binary in the calculation of the $S_2$ gives us identical results. This can be easily seen when comparing Eqs. (8) and (10). If we assume that in the rdfs we can replace 3 by 1 and 4 by 2, then Eq. (9) can be rewritten as follows:

$$\begin{aligned}\frac{S_2^Q}{k_B} &= -2\pi\rho \int_0^\infty r^2 dr \\ &\times \Big[ (\chi_1'\chi_1' + 2\chi_1'\chi_3' + \chi_3'\chi_3')\{g_{11}(r)\ln g_{11}(r) - (g_{11}(r) - 1)\} \\ &+ (\chi_1'\chi_2' + \chi_1'\chi_4' + \chi_3'\chi_2' + \chi_3'\chi_4') \\ &\times \{g_{12}(r)\ln g_{12}(r) - (g_{12}(r) - 1)\} \\ &+ (\chi_2'\chi_1' + \chi_2'\chi_3' + \chi_4'\chi_1' + \chi_4'\chi_3') \\ &\times \{g_{21}(r)\ln g_{21}(r) - (g_{21}(r) - 1)\} \\ &+ (\chi_2'\chi_2' + 2\chi_2'\chi_4' + \chi_4'\chi_4')\{g_{22}(r)\ln g_{22}(r) - (g_{22}(r) - 1)\} \Big] \\ &= -2\pi\rho \int_0^\infty r^2 dr \Big[ (\chi_1' + \chi_3')^2 \{g_{11}(r)\ln g_{11}(r) - (g_{11}(r) - 1)\} \\ &+ (\chi_1' + \chi_3')(\chi_2' + \chi_4')\{g_{12}(r)\ln g_{12}(r) - (g_{12}(r) - 1)\} \\ &+ (\chi_2' + \chi_4')(\chi_1' + \chi_3')\{g_{21}(r)\ln g_{21}(r) - (g_{21}(r) - 1)\} \\ &+ (\chi_2' + \chi_4')^2 \{g_{22}(r)\ln g_{22}(r) - (g_{22}(r) - 1)\} \Big] \\ &= -2\pi\rho \int_0^\infty r^2 dr \sum_{i,j=1}^2 \chi_i\chi_j\{g_{ij}(r)\ln g_{ij}(r) - (g_{ij}(r) - 1)\}. \quad (C1)\end{aligned}$$

The last equality can be written because $\chi_1 = \chi_1' + \chi_3'$ and $\chi_2 = \chi_2' + \chi_4'$.

In Fig. 11, we plot $S_2^M$, where the change in structure due to the pinned particles is considered. We find that, similar to unpinned systems, at high temperatures, $S_2^M$ is larger than $S_{ex}^M$, and at low temperatures, the scenario is reversed. This leads to a crossover between $S_2^M$ and $S_{ex}^M$. In earlier studies, we have shown that this cross over point is close to the onset temperature.[49,67] Here, we find that with increased pinning density, similar to the onset temperature, the cross over point moves to higher temperatures. It is important to note that for a single set of simulations, the pinning positions remain constant over time. Thus, the pin–pin partial rdf is noisy and cannot be used to calculate the structural parameters where we need to integrate the rdf. The comparison plot of the $g_{33}$ averaged over $10^4$ different initial pinned configurations, and that from a single configuration is plotted in Fig. 12. We utilize this averaged partial rdfs for the calculation of $S_2$ in Fig. 11.

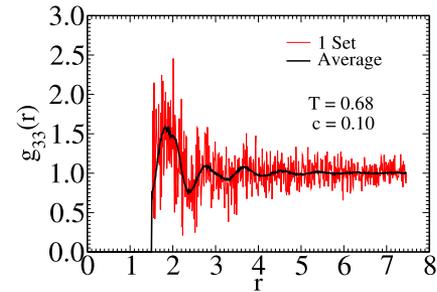

**FIG. 12.** Radial distribution function (rdf) between two pinned A particles, $g_{33}(r)$. The black line represents the average radial distribution function (RDF) obtained from $10^4$ different initial pinned configurations. The red line represents the date from a single set.

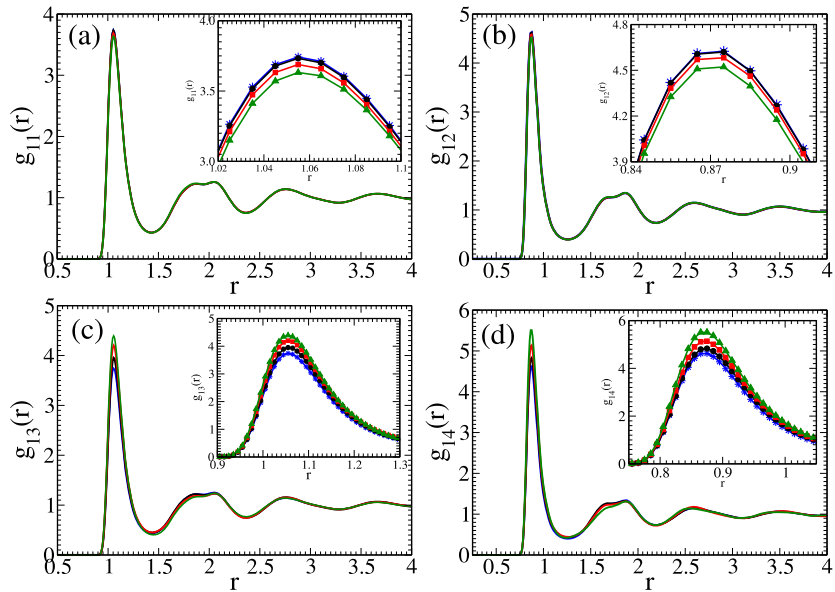

**FIG. 13.** Radial distribution function, g(r), while treating the pinned system as a quaternary system at T = 0.68. (a) $g_{11}$ as a function of r. (b) $g_{12}$ as a function of r. (c) $g_{13}$ as a function of r. (d) $g_{14}$ as a function of r. Inset: We have zoomed onto the first peak of the respective figures, which clearly shows the difference in the radial distribution functions. Note that color coding is similar to Fig. 1. Here, we refer to mobile A type of particles as 1, mobile B type of particles as 2, pinned A type of particles as 3, and pinned B type of particles as 4.

The temperature where these two entropies cross each other is the $\Delta S = 0$ [Eq. (7)] point. For the KA model (c = 0) and other systems, it was earlier shown that the temperature where these two entropies cross is similar to the onset temperature of glassy dynamics.[49,67] However, it has also been found that in systems with mean field like characteristics, the temperature where $\Delta S = 0$ is lower than the onset temperature.[69,70] The latter scenario is similar to what we find for pinned systems. We find that with the increase in pinning, the difference between the onset temperature and the temperature where the two entropies cross increases. Interestingly, a similar difference between the freezing point and $\Delta S = 0$ was observed for higher dimensional systems[71] and the Gaussian core model (GCM).[72] Note that if the pair excess entropy is calculated assuming the pinned system to be a binary system, then the cross over between the pair excess entropy and the total entropy will take place at unphysically low temperatures.

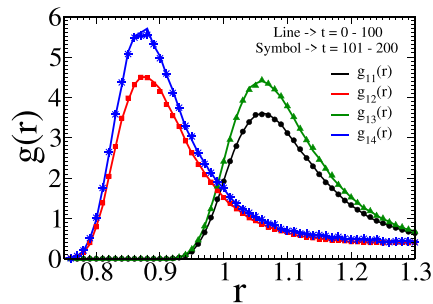

**FIG. 14.** Radial distribution function, g(r) at different time intervals for c = 0.15 system. The solid line and symbol represent the radial distribution function at times $t = 0$–100 and $t = 100$–200, respectively.

## APPENDIX D: RADIAL DISTRIBUTION FUNCTION

In Fig. 13 (assuming the pinned particles are of a different species), we find that with increased pinning density, the partial rdfs start deviating from the c = 0 system. With an increase in "c," there is a drop in the peak value of the rdfs between two mobile particles ($g_{11}$, $g_{12}$). On the other hand, the height of the first peak of the partial rdfs between mobile and pinned particles ($g_{13}$, $g_{14}$) grows with "c."

We observe that this structural change happens quickly, immediately after the pinning process. In Fig. 14, we plot $g(r)$, averaged from $t = 0$–100 and $t = 101$–200, where the pinning is done at t = 0. We find that both rdfs overlap. This is shown for the c = 0.15 system, where the difference is the maximum.

We also show that $\chi'_1 g_{11} + \chi'_3 g_{13}$ is the same as $\chi_A g_{AA}$, and $\chi'_2 g_{12} + \chi'_4 g_{14}$ is the same as $\chi_B g_{BB}$ (Fig. 15). This is precisely why we do not see a change in structure when the pinned particles are not treated as a different species (Fig. 1).

To check the system size dependence, in Fig. 16, we plot the rdfs for a 4000 particle and a 1000 particle system. We find that the change in the rdf with pinning is almost independent of the system size, with the difference between the rdfs of the unpinned and pinned systems increasing marginally for larger system sizes.

We also check the dependence of the rdf on the integration time $\Delta t$ (Fig. 17). From this plot, we observe that the rdf is independent of the integration time step.

## APPENDIX E: ISOCONFIGURATION RUN (IC)

To quantify the dependence of the dynamics on the structure and particle size, we perform isoconfigurational runs (IC). IC is a powerful technique introduced by Harrowell and co-workers to investigate the role of structure in the dynamical heterogeneity of particles.[52–55] Among other factors, a particle's displacement can depend on its structure and also on its initial momenta. This technique was proposed to remove the uninteresting variation in the particle displacements arising from the choice of initial momenta and extract the role of the initial configuration on the dynamics and its heterogeneity. For each system, five different isoconfigurational runs are carried out for 4000 particles. To ensure that all configurations are different, the configurations are chosen such that the two sets are greater than $100\tau_\alpha$ apart. All five ICs have different structures as well as different pin particle positions. Note that since we have shown in Sec. III D that after pinning, the structure changes, after we pin the equilibrium position of the mobile particles, we run the system for t = 100 timesteps and then consider that as our initial configuration. We run 100 trajectories for each configuration, with different starting velocities randomly assigned

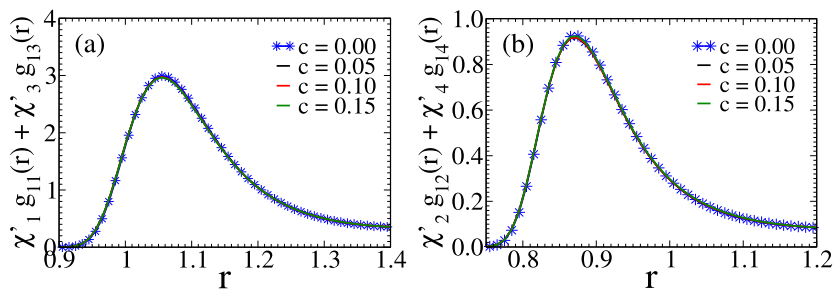

**FIG. 15.** (a) $(\chi'_1 g_{11} + \chi'_3 g_{13})$ of the pinned system and $\chi_A g_{AA}$ of the KA system as a function of r. (b) $(\chi'_2 g_{12} + \chi'_4 g_{14})$ of the pinned system and $\chi_B g_{AB}$ of the KA system as a function of r.

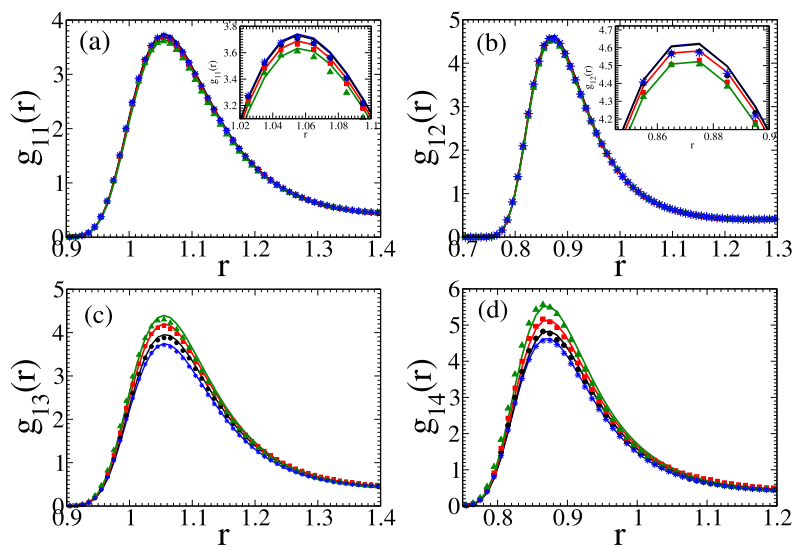

**FIG. 16.** System size dependence in radial distribution function, g(r) while treating the pinned system as a quaternary system, at T = 0.68. (a) $g_{11}$ as a function of r. (b) $g_{12}$ as a function of r. (c) $g_{13}$ as a function of r. (d) $g_{14}$ as a function of r. Inset: We have zoomed onto the first peak of the respective figures, which clearly shows the difference in the radial distribution functions. Note that color coding is similar to Fig. 1. Here, we refer to mobile A type of particles as 1, mobile B type of particles as 2, pinned A type of particles as 3, and pinned B type of particles as 4. The solid line represents the 4000 particle system, and the symbol represents the 1000 particle system.

from the Maxwell–Boltzmann distribution for the corresponding temperatures.

Mobility, μ, is the average displacement of each particle over these 100 runs and is calculated as follows:[54]

$$\mu^j(t) = \frac{1}{N_{IC}} \sum_{i=1}^{N_{IC}} \sqrt{\left(r_i^j(t) - r_i^j(0)\right)^2}, \quad \text{(E1)}$$

where the jth particle's mobility at time t is represented by the term $\mu^j(t)$. The position of the jth particle in the ith trajectory at time t is denoted by the term $r_i^j(t)$, and its initial position is denoted by the term $r_i^j(0)$. The sum of the values is calculated over each of the $N_{IC}$ trajectories that were carried out during the isoconfiguration runs. We determine the average displacement or mobility for the $j$th particle at time $t$ by averaging these displacements over all trajectories.

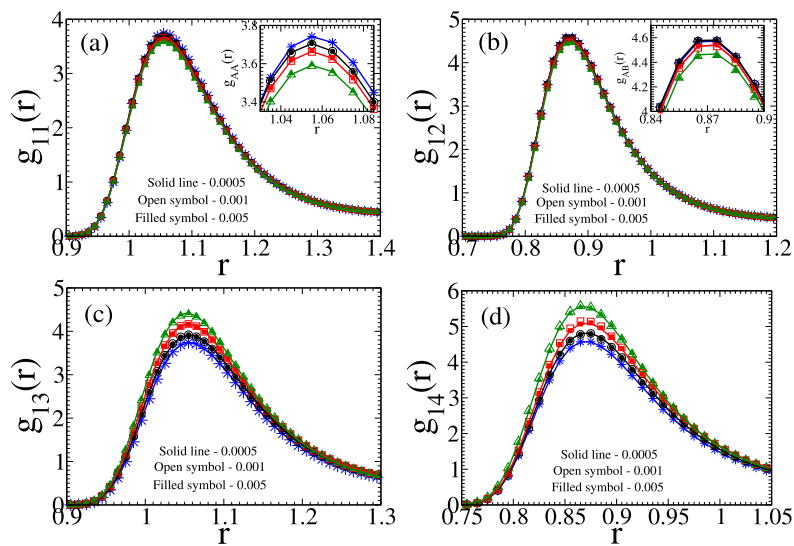

**FIG. 17.** Effect of the integration time step, Δt, on radial distribution function, g(r) while treating the pinned system as a quaternary system at T = 0.68. (a) $g_{11}$ as a function of r. (b) $g_{12}$ as a function of r. (c) $g_{13}$ as a function of r. (d) $g_{14}$ as a function of r. Inset: We have zoomed onto the first peak of the respective figures, which clearly shows the difference in the radial distribution functions. Note that color coding is similar to Fig. 1. Here, we refer to mobile A type of particles as 1, mobile B type of particles as 2, pinned A type of particles as 3, and pinned B type of particles as 4.

## APPENDIX F: IDENTIFICATION OF FAST PARTICLES

There are various methods available for identifying fast particles in the literature.[73–77] In our study, we employ the approach proposed by Candelier et al.[76,77] This method involves the calculation of a quantity called $p_{hop}(i, t)$ for each particle within a specified time window $W = [t_1, t_2]$.

The $p_{hop}(i, t)$ quantity captures the rate of change in the average position of a particle, indicating the occurrence of a cage jump. The expression for $p_{hop}(i, t)$ is given as follows:[78]

$$p_{hop}(i,t) = \sqrt{\langle (r_i - \langle r_i \rangle_U)^2 \rangle_V \langle (r_i - \langle r_i \rangle_V)^2 \rangle_U}, \quad (F1)$$

where $r_i$ represents the position of particle $i$, and $\langle \cdot \rangle$ denotes the averages over time. The time window $W$ is divided into two intervals, $U = [t - \Delta t/2, t]$ and $V = [t, t + \Delta t/2]$. By calculating $p_{hop}(i, t)$ for each particle, we can determine whether a particle experiences a significant change in its average position, indicating its involvement in cage jumps and enhanced dynamics. In our analysis, we compare the calculated $p_{hop}(i, t)$ values to a threshold value $p_c$, which is determined as the mean square displacement, $\langle \Delta r(t)^2 \rangle$ at a specific time $t_{max}$ where the non-Gaussian parameter, $\alpha_2 = \frac{3\langle \Delta r(t)^4 \rangle}{(\langle \Delta r(t)^2 \rangle)^2} - 1$ is maximized. If $p_{hop}(i, t)$ exceeds $p_c$, we identify the particle as a fast particle.[33,67,79]

It is important to note that in our study, we specifically analyzed the structure and dynamics of mobile A particles. Therefore, we calculate the Mean Square Displacement (MSD) and the non-Gaussian parameter specifically for mobile A particles. For a more comprehensive understanding of the method and its application in our study, we refer readers to Refs. 33, 67, 78, and 79.

## APPENDIX G: CORRELATION BETWEEN STRUCTURE AND DYNAMICS AND PREDICTION OF ONSET TEMPERATURE

In this section, we use the structure dynamics correlation to identify the onset temperature of the glassy dynamics, a methodology used in earlier studies.[33,78]

We identify fast particles using a well-documented method[33,76,77] (details are given in Appendix F). In Fig. 18, we plot $P_R(1/\beta\Phi_r)$ as a function of temperature for different $1/\beta\Phi_r$ values and find that it can be expressed in an Arrhenius form, $P_R(1/\beta\Phi_r) = P_0(1/\beta\Phi_r) \exp(\Delta E(1/\beta\Phi_r)/T)$, where activation energy, $\Delta E$, is a function of $1/\beta\Phi_r$ and is higher for smaller $1/\beta\Phi_r$ values. The plots cross at a certain temperature, which describes the limiting temperature where the theory is valid[33] and has been identified earlier as the onset temperature of the supercooled liquid.[33,50,78]

In this analysis, we find that when we treat the system as binary, the onset temperature remains similar for all pinning concentrations. However, when we treat the system as a modified quaternary system, the onset temperature increases with increasing pinning concentration.[19] As we show in Appendix A, this predicted onset temperature is similar to the onset temperature predicted from the well-known inherent structure energy method (Fig. 8 and Table I).[56]

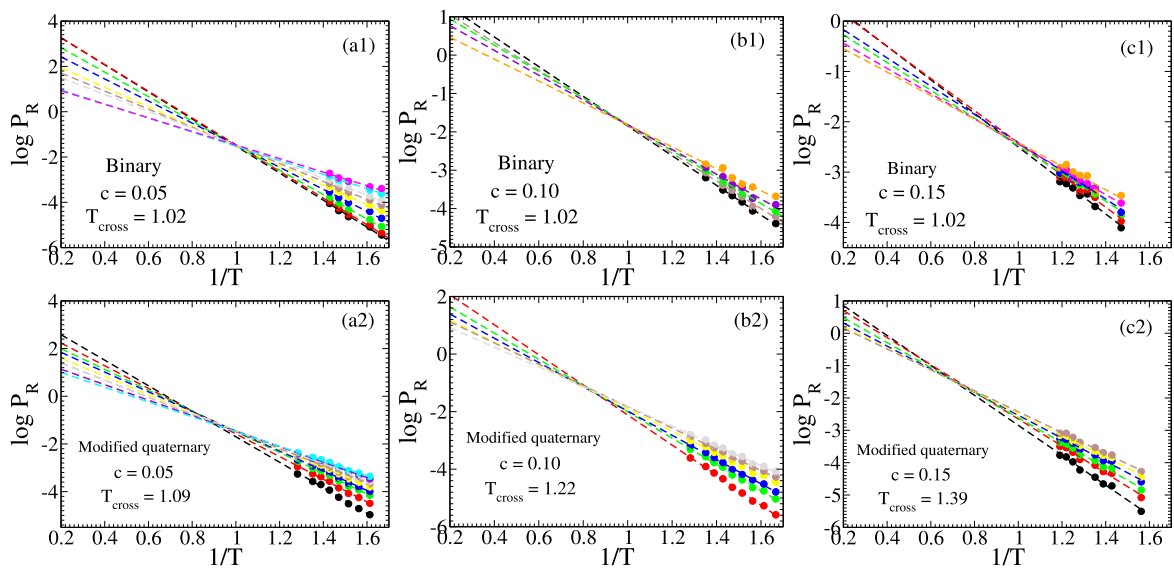

**FIG. 18.** log $P_R(1/\beta\Phi_r)$ as a function of 1/T at different values of the SOP $(1/\beta\Phi_r)$. Top panel: In the calculation of the SOP, the pinned system is treated as a binary system. (a1) c = 0.05, (b1) c = 0.10, and (c1) c = 0.15. Bottom panel: In the calculation of the SOP, the pinned system is treated as a modified quaternary system. (a2) 0.05, (b2) c = 0.10, and (c2) c = 0.15.